\documentclass[runningheads]{llncs}
\usepackage{amsmath}
\usepackage{amssymb}
\usepackage{enumerate}
\usepackage{graphicx}
\usepackage{array}

\newcolumntype{M}[1]{>{\centering\arraybackslash}m{#1}}
\newcolumntype{N}{@{}m{0pt}@{}}
\newcolumntype{C}{ >{\centering\arraybackslash} m{1.9cm} }
\newcolumntype{D}{ >{\centering\arraybackslash} m{3cm} }
\newcolumntype{H}{ >{\centering\arraybackslash} m{3.4cm} }
\newcolumntype{W}{ >{\centering\arraybackslash} m{1.1cm} }
\newcolumntype{R}{ >{\centering\arraybackslash} m{1.2cm} }

\begin{document}
\title{A Joint Encryption-Encoding Scheme Using QC-LDPC Codes Based on Finite Geometry\thanks{This work was partially supported by Iran National Science Foundation (INSF) under contract No. 92.32575.}}
\titlerunning{A Joint Enc-Enc based on FG-QC-LDPC}
%
\author{Hossein Khayami\inst{1}\orcidID{0000-0002-4485-9322} \and
Taraneh Eghlidos\inst{2}\orcidID{0000-0002-3182-0277} \and
Mohammad Reza Aref\inst{1}}
\authorrunning{H. Khayami et al.}
%
\institute{Information Systems and Security Laboratory, Department of Electrical Engineering, Sharif University of Technology, Tehran 11155-11365, Iran \and
Electronics Research Institute, Sharif University of Technology, Tehran 11155-11365, Iran\\}
\maketitle              
\begin{abstract}
Joint encryption-encoding schemes have been released to fulfill both reliability and security desires in a single step. Using Low Density Parity Check (LDPC) codes in joint encryption-encoding schemes, as an alternative to classical linear codes, would shorten the key size as well as improving error correction capability. In this article, we present a joint encryption-encoding scheme using Quasi Cyclic-Low Density Parity Check (QC-LDPC) codes based on finite geometry. We observed that our proposed scheme not only outperforms its predecessors in key size and transmission rate, but also remains secure against all known cryptanalyses of code-based secret key cryptosystems. We subsequently show that our scheme benefits from low computational complexity. In our proposed joint encryption-encoding scheme, by taking the advantage of QC-LDPC codes based on finite geometries, the key size decreases to 1/5 of that of the so far best similar system. In addition, using our proposed scheme a wide range of desirable transmission rates are achievable. This variety of codes makes our cryptosystem suitable for a number of different communication and cryptographic standards.

\keywords{Joint encryption-encoding \and secure channel coding \and QC-LDPC \and code-based cryptography \and finite geometry.}
\end{abstract}
\section{Introduction}\label{S1Intro}
The first code-based cryptosystem has been introduced by McEliece \cite{mceliece1978public}. The security of this cryptosystem is based on the general decoding problem, which is known to be NP-complete problem \cite{Berlekamp1978}. Although at the time of writing this paper, no algorithm running on quantum computers has been published to break the code-based cryptosystems, its large key size and low transmission rate in comparison with the prevalent cryptosystems such as RSA and ElGamal made these cryptosystems unusable from implementation and standard perspectives.

After McEliece published his public key code-based cryptosystem \cite{Berlekamp1978}, Rao in 1984 proposed a secret key cryptosystem based on the McEliece public key cryptosystem \cite{Rao1984}. In 1986, Rao and Nam made a security modification \cite{RaoNam1986Crypto}. In 1987, Struik and Tilburg pointed out some weaknesses of the Rao-Nam cryptosystem and proposed an improved version \cite{StruikTilburg1987}. In 2000, a secret key code-based cryptosystem with much shorter key was introduced \cite{Barbero2000}.

In conventional communication systems the encryption and encoding is being performed separately and in series. In 2008, a joint encryption-encoding scheme has been proposed with lower complexity in comparison with separate encryption and encoding has been proposed \cite{SobhiAfshar2009}. This scheme, using quasi-cyclic structure, succeeded in shortening the key size. Moreover, it gains benefits from the fast decoding algorithms and superior error performance of the LDPC codes. In 2012, another scheme using QC-LDPC codes based on Extended Difference Families (EDF) was proposed \cite{hooshmand2015improving}, which could not achieve further improvement on the key size. In 2014, a joint encryption-encoding scheme with the novel idea of puncturing, instead of adding a perturbation vector, was introduced \cite{Esmaeili2014}. Then, in 2015, Esmaeili and Gulliver added random insertions to improve the security of their system \cite{Esmaeili2015}. In \cite{Esmaeili2016}, they added an agreed random error vector to their encryption process. In \cite{Esmaeili2017}, they used random interleaving instead of random insertions and deletions. The cryptosystem in \cite{Esmaeili2016,Esmaeili2017} is an encoding then encryption system rather than a joint encryption-encoding scheme. Besides, it is shown that the use of two pairs of LFSRs has made Esmaeili-Gulliver cryptosystem vulnerable against ciphertext-only attack \cite{Lee2017}. A recent work on joint encryption-encoding, which uses polar codes as generator matrix, claimed to achieve a relatively smaller key size, higher level of security, and higher code rate \cite{mafakheri2017}.

Although the key sizes of recent schemes reduced considerably in comparison to the trailblazing code-based studies, the shortest of them is about 13 times larger than the key size of conventional AES-128. Due to this fact, attaining a more compact secret key is one of our motivations of this article. Besides shortening the secret key, increasing the transmission rate, decreasing the computational complexity of algorithm, and efficiently correcting channel errors, as well as keeping the cryptosystem secure are the most challenging issues in joint encryption-encoding researches. Solving these issues need a proper family of codes to be utilized. This code should possess the following characteristics:
\begin{itemize}
	\item Efficiently decodable
	\item A family with large cardinality
	\item Achievable high transmission rate
	\item Compressible matrices in the secret key
\end{itemize}

In this paper, we propose a joint encryption-encoding scheme utilizing QC-LDPC codes based on finite geometry (FG-QC-LDPC) in order to obtain our goal, which is introducing a practical solution for the above issues. In finite geometry, every line can be identified through each of two points lying on that. This property enables shortening the key size to 1/5 of the so far best known joint encryption-encoding scheme. The wide acceptable range in the parameters of our proposed scheme makes it suitable for various applications and different levels of security. Furthermore, we show that the FG-QC-LDPC joint scheme is secure against all known cryptanalyses of such schemes.

The rest of this paper is organized as follows. Section \ref{S2Pre} recalls some basic facts about finite geometry and QC-LDPC codes derived from them. Next, the description of our new joint encryption-encoding scheme using FG-QC-LDPC codes is given in Section \ref{S3scheme}. The security and performance including key size, error performance, and complexity of our scheme are discussed in Section \ref{S4Sec}. Finally, Section \ref{S5con} summarizes and concludes the paper.

\section{Preliminaries}\label{S2Pre}
We took the advantages of FG-QC-LDPC codes to achieve our designated goals, namely improving the performance in comparison to the best known systems in the literature and also keeping the system secure against all known cryptanalyses. The basic definitions of QC-LDPC codes based on finite geometry is summarized in this section.

\subsection{QC-LDPC}
In cryptographic applications, quasi-cyclic LDPC codes allow us to reduce the key size as well as the complexity in comparison with the general LDPC codes \cite{Baldi2007c}. The parity check matrix of a QC-LDPC code is represented as follows,
\begin{equation}\label{H_QC}
H=
\left[ \begin{array}{@{}*{3}{c}@{}}
H_{0,0} & \dots  & H_{0,n_0-1} \\
\vdots & \ddots & \vdots \\
H_{r_0-1,0} & \dots  & H_{r_0-1,n_0-1}
\end{array}\right]
\end{equation}
where each $H_i$ is a circulant block of size $p\times p$. 

There are different families of QC-LDPC codes used in code-based cryptography, namely, the Extended Difference Family (EDF) \cite{Baldi2007c,hooshmand2015improving}, and the Random Difference Family (RDF) \cite{Baldi2007a,Baldi2013improving}. In our scheme we propose using finite geometry to construct circulant blocks of the parity check matrix. This helps us to attain a shorter secret key than which is available in the literature for joint encryption-encoding schemes.

\subsection{Finite Geometry}
A finite geometry is composed of finite number of points. In this paper we focus on two types of finite geometry, namely projective geometry (PG) and Euclidean geometry (EG). The definitions of this subsection are generally provided from \cite{lin2004error} and \cite{Ryan2009}.

\subsubsection{Euclidean Geometry}
\begin{definition}\label{def1}
	(Euclidean geometry). All $m$-tuples $(a_0,a_1,\dots,a_{m-1})$ with $a_i$ from $GF(q=p^s)$ where $p$ is prime and $s$ is a natural number form a vector space. This vector space is also known as the finite Euclidean geometry of dimension $m$ over $GF(q)$, denoted by $EG(m,q)$.
	Vector additions and scalar multiplications of these $m$-tuples are conducted in $GF(q)$.
\end{definition}
\begin{definition}\label{def2}
	(point). Each $m$-tuple $\mathbf{a}=(a_0,a_1,\dots,a_{m-1})$ represents a point in $EG(m,q)$.
\end{definition}

\begin{definition}\label{def3}
	(origin). The all-zero $m$-tuple $\mathbf{0}=(0,0,\dots,0)$ is called the origin.
\end{definition}

\begin{definition}\label{def4}
	(line). The set of $\{\mathbf{a_0}+\beta \mathbf{a}|\beta \in GF(q),\mathbf{a}\neq\mathbf{0}\}$ is a line, which is composed of $q$ points.
\end{definition}
The number of points in $EG(m,q)$ is equal to the number of all $m$-tuples i.e. $n=q^m$. For every two distinct points there exists exactly one line connecting them. The number of lines intersecting at each point can be obtained by dividing the number of possible second points of that line by the number of other points in each line.
\begin{equation}
\gamma =\frac{q^m-1}{q-1}
\end{equation}
Therefore, the number of lines in the $EG(m,q)$ is as follows,
\begin{equation}
J=q^{m-1}\frac{q^m-1}{q-1}
\end{equation}
\subsubsection{Euclidean Geometry Without Origin}
By omitting the origin and all lines intersecting at the origin, a new geometry appears which is denoted by $EG^*(m,q)$.

If $\alpha$ is primitive in $GF(q^m)$, then $\alpha^i$ for $0\leq i \leq q^m-2$ represents the elements of $GF(q^m)$. So the incident vector of line $F$ is as given below,
\begin{equation}
V_F=(v_0,v_1,\dots,v_{q^m-2})
\end{equation}
where $v_i=1$ if the line $F$ intersects at point $\alpha^i$, otherwise $v_i=0$.

In this geometry, a circularly shifted incident vector of a line is an incident vector for another line \cite{Ryan2009}. This property partitions the set of all lines,  $J_o$,  into $\mathcal{N}_{c,EG^*}$ cyclic classes;
\begin{equation}
\mathcal{N}_{c,EG^*}=\frac{J_o}{n}=\frac{q^{m-1}-1}{q-1}.
\end{equation}

These cyclic classes enable us to generate circulant blocks for parity check matrices of QC-LDPC codes. All the necessary information of Euclidean geometry is summarized in Table \ref{table_EG}.

\begin{table}[!t]
	\renewcommand{\arraystretch}{2}
	\caption{Parameters of the Euclidean Geometry}
	\label{table_EG}
	\centering
	
	\begin{tabular}{|M{4.5cm}||D|H|}
		\hline
		\bfseries \centering Parameters & \bfseries $\mathbf{EG(m,q)}$ & \bfseries  $\mathbf{EG^*(m,q)}$\\[3pt]
		\hline\hline
		\bfseries Field & $GF(q)$ & $GF(q)$  \\
		\bfseries  Dimension & $m$ & $m$  \\
		\bfseries  No. of points & $n=q^m$ & $n=q^m-1$  \\
		\bfseries  No. of lines & $J=q^{m-1}\frac{q^m-1}{q-1}$ & $J_o=\frac{(q^{m-1}-1)(q^m-1)}{q-1}$  \\
		\bfseries  No. of points in each line & $\rho=q$ & $\rho=q$  \\ 
		\bfseries  No. of lines intersecting at each point & $\gamma =\frac{q^m-1}{q-1}$ & $\gamma =\frac{q^m-1}{q-1}-1$  \\
		\bfseries  No. of cyclic classes & ----- & $\mathcal{N}_{c,EG^*}=\frac{J_o}{n}=\frac{q^{m-1}-1}{q-1}$  \\
		\hline
	\end{tabular}
\end{table}

\subsubsection{Projective Geometry}
Consider the Galois field $GF(q^{m+1})$ and $\alpha$, a primitive element in this field. So $\alpha^0,\alpha^1,\dots,\alpha^{q^{m+1}-2}$ constitute all non-zero elements of $GF(q^{m+1})$. Let $n=\frac{q^{m+1}-1}{q-1}$ and $\beta=\alpha^n$. Then the order of $\alpha$ is $q-1$. Now, $0,\beta^0,\beta^1,…,\beta^{q-2}$ form the elements of $GF(q)$. Considering the definition of $\alpha$ and $\beta$, all non-zero elements of $GF(q^{m+1})$ can be partitioned into $n$ disjoint subsets as shown below,
\setlength{\arraycolsep}{0.0em}
\begin{eqnarray}\label{alpha_beta}
&(\alpha^0)\triangleq \{ \alpha^0,\beta\alpha^0,\beta^2\alpha^0,\dots,\,\beta^{q-2}\alpha^0 \} \nonumber \\ 
&(\alpha^1)\triangleq \{ \alpha^1,\beta\alpha^1,\beta^2\alpha^1,\dots,\,\beta^{q-2}\alpha^1 \} \nonumber \\ 
& {}\:\vdots  \nonumber \\
&(\alpha^{n-1})\triangleq \{ \alpha^{n-1},\beta\alpha^{n-1},\beta^2\alpha^{n-1},\dots,\,\beta^{q-2}\alpha^{n-1} \}
\end{eqnarray}
\setlength{\arraycolsep}{5pt}
Each of the above subsets represent a distinct point in projective geometry, which is denoted by $PG(m,q)$. In this geometry, each line consists of $q+1$ points, formed by linear combination of two distinct $\alpha^{j_1}$ and $\alpha^{j_2}$ points;
\begin{eqnarray}\label{line_def}
(\eta_1 \alpha^{j_1} + \eta_2 \alpha^{j_2})  & ; &  \eta_i \in GF(q).
\end{eqnarray}
The number of lines intersecting at every particular point is $\gamma = \frac{n-1}{q} = \frac{q^m-1}{q-1}$, which is obtained by dividing the remaining number of points chosen as the second point of line $(=n-1)$ by the number of other points in each line $(=q)$.

Let $F=(\eta_1 \alpha^{j_1} + \eta_2 \alpha^{j_2});\eta_i \in GF(q)$ be a line in $PG(m,q)$, then for all $0\leq i < n$, $\alpha^i F$ is also a line in $PG(m,q)$. The $\alpha^i F$ is called the $i^{th}$ circular shift of line $F$.

If $m$ is even, all lines in $PG(m,q)$ have primitive incident vector and partitioned into $\mathcal{N}_{c,even}=  \frac{q^m-1}{q^2-1}$ cyclic classes, where each cyclic class consists of $n$ lines. If $m$ is odd, only $J_0$ lines of $PG(m,q)$ have primitive incident vector \cite{Ryan2009}.
\begin{eqnarray}\label{PG_cyclic_class}
J_o=\frac{q(q^{m+1}-1)(q^{m-1}-1)}{(q^2-1)(q-1)}
\end{eqnarray}
These incident vectors are partitioned into $\mathcal{N}_{c,odd}$ cyclic classes.
\begin{eqnarray}\label{N_C_odd}
\mathcal{N}_{c,odd}=\frac{q(q^{m-1}-1)}{(q^2-1)}
\end{eqnarray}
Table \ref{table_PG} summarizes necessary information of the projective geometry.
\begin{table}[!t]
	\renewcommand{\arraystretch}{2}
	\caption{Parameters of the Projective Geometry}
	\label{table_PG}
	\centering
	
	\begin{tabular}{|M{7cm}||H|}
		\hline
		\bfseries \centering Parameters & \bfseries Value \\[3pt]
		\hline\hline
		\bfseries Field & $GF(q)$  \\
		\bfseries  Dimension & $m$ \\
		\bfseries  No. of field's elements that consist each point & $q-1$  \\
		\bfseries  No. of points & $n=\frac{q^{m+1}-1}{q-1}$  \\
		\bfseries  No. of lines & $J=\frac{n\gamma}{\rho}=\frac{(q^{m+1}-1)(q^m-1)}{(q-1)(q-1)(q+1)}$  \\
		\bfseries  No. of points in each line & $\rho=q+1$   \\ 
		\bfseries  No. of lines intersecting at each point & $\gamma =\frac{q^m-1}{q-1}$  \\
		\bfseries  No. of cyclic classes (even m) & $\mathcal{N}_{c,even}=\frac{q^{m}-1}{q^2-1}$  \\
		\bfseries  No. of cyclic classes (odd m) & $\mathcal{N}_{c,odd}=\frac{q(q^{m-1}-1)}{q^2-1}$  \\
		\hline
	\end{tabular}
\end{table}

In finite geometry, since there is exactly one line connecting two distinct points, no two incident vectors have more than one non-zero elements in the same location. As a result of this property, the girth of QC-LDPC codes based on finite geometry is at least 6.
\subsection{FG-QC-LDPC Codes}
In our scheme we exploit a QC-LDPC code with one block row of the form
\begin{eqnarray}\label{H_QC_oneblockrow}
H_{qc}=[H_0 H_1 \dots H_{n_0-1}].
\end{eqnarray}
In FG-QC-LDPC codes, as a subset of QC-LDPC codes, the first rows of the circulant blocks are derived from incident vectors of a line in that geometry. Thanks to the geometric construction, each line, and therefore its incident vector, can be identified by only two points lying on that line. This property helps us to shorten the key size. Other details regarding the key size are mentioned in Section \ref{S4Sec}.

The number of circulant blocks in the parity check matrix, $n_0$, is limited to the number of cyclic classes in that particular geometry, i.e. $n_0 \leq \mathcal{N}_c$. The parity check matrix derived from finite geometry has the following characteristics, which correspond to the parameters given in Tables \ref{table_EG} and \ref{table_PG}.
\begin{itemize}
	\item 	The Hamming weight of each row in each circulant block is equal to the number of points lying on each line in that finite geometry.
	\item	The size of each circulant block is $p \times p$, where $p$ is the total number of points in that geometry.
	\item	No two columns have more than one common locations of ‘1’s. This is due to the fact that two distinct lines in finite geometry are either disjoint or intersect at only one point.
	\item	The Tanner graph contains no length 4 cycles.
	\item   The codeword length is $n=n_0 \times p$.
	\item 	The length of message vectors or equally the dimension of the code is $k=(n_0-1)\times p = k_0 \times p$.
\end{itemize}
\section{FG-QC-LDPC Joint Encryption-Encoding Scheme}\label{S3scheme}
Here is the description of the proposed joint encryption-encoding scheme based on FG-QC-LDPC codes in three different steps, that is, key generation, encryption-encoding, and decryption-decoding. Then, we discuss the range of suitable parameters for our proposed scheme.
\subsection{Key Generation}\label{Key_gen}
The secret key of the joint encryption-encoding scheme is composed of a parity check matrix, $H$, a permutation matrix, $P$, and the seed of the Pseudo Random Number Generator (PRNG). 
\subsubsection{Parity Check Matrix}
The parity check matrix of the FG-QC-LDPC code in Eq.(\ref{H_QC_oneblockrow}) can be constructed based on either Euclidean geometry or projective geometry. The construction procedure first start with choosing between these two types of geometries and their parameters. According to Section \ref{S2Pre}, a finite geometry is defined in terms of two parameters, that is, its dimension, $m$, and the corresponding field, $GF(q)$.

As discussed in Section \ref{S2Pre}, in both cases of Euclidean and projective geometries all lines are partitioned into different cyclic classes. Each cyclic class in a finite geometry forms a set of rows of a circulant block. Thus, for generating a circulant block $H_i$ one needs to simply specify a cyclic class and then assign only its first row.

In the case of EG, the number of cyclic classes, according to the Table \ref{table_EG} is $\mathcal{N}_{c,EG^*}=\frac{J_o}{n}=\frac{q^{m-1}-1}{q-1}$ and the number of lines in each cyclic class is $p=q^m-1$ which is equal to the length of each row vector of circulant blocks.

In the case of PG, the number of cyclic classes is  $\mathcal{N}_{c,even}=\frac{q^{m}-1}{q^2-1}$ or $\mathcal{N}_{c,odd}=\frac{q(q^{m-1}-1)}{q^2-1}$, when the dimension of the geometry is even or odd, respectively. Here, the number of lines in each cyclic class is $p=\frac{q^{m+1}-1}{q-1}$.

To sum up with the generation of the parity check matrix, we should choose public parameters, that is, the type of geometry, its dimension m, its corresponding field $GF(q)$, and the number of circulant blocks of the matrix, $n_0$. Then each circulant block must be generated in the above fashion.

\subsubsection{Permutation Matrix}
In our scheme, the permutation matrix is a block diagonal matrix in the form of
\begin{eqnarray}\label{P}
P=
\begin{pmatrix}
\pi & \dots & 0 \\
\vdots & \ddots & \vdots \\
0 & \dots & \pi
\end{pmatrix}_{n_0l \times n_0l},
\end{eqnarray}
where each $\pi$ block is an $l \times l$ permutation matrix.
\subsubsection{The PRNG Seed}
In order to generate a sequence of perturbation vectors $e_P$ we should utilize a PRNG. Thus, in order to use the same sequence as perturbation vectors by the transmitter and the receiver, it suffices they agree on the same seed for the PRNG. The sequence generated by the PRNG is then divided into $(n-k)$-bit vectors, $z$. The perturbation vectors are computed by $e_P=H^{-1}z$, where $H^{-1}$ is the right inverse of $H$. Therefore, the perturbation vector $e_P$ is of length $n$. Different PRNGs can be employed depending on the hardware/software resources and applications of the joint encryption-encoding scheme.
\subsection{Encryption-Encoding}
For doing joint encryption-encoding, the transmitter needs to compute the generator matrix $G$ from the parity check matrix $H$. In QC-LDPC codes with a parity check matrix in the form of one block row (see Eq.(\ref{H_QC_oneblockrow})), the generator matrix can be constructed as given below,
\begin{eqnarray}\label{H_G}
G=
\left(\begin{matrix} I_k  \end{matrix}\middle| \begin{matrix} 
{( H_{n_0-1}^{-1}H_0)}^T\\ \vdots \\ {( H_{n_0-1}^{-1}H_{n_0-2})}^T
\end{matrix}\right).
\end{eqnarray}
Note that for $G$ being used as the generator matrix, it is sufficient for at least one circulant block, $H_i$, to be non-singular. Without loss of generality, we assume that the circulant block $H_{n_0-1}$ is a non-singular matrix. 

Next, the transmitter generates a perturbation vector $e_P$ as given below,
\begin{eqnarray}
e_P=H^{-1} z,
\end{eqnarray}
where $z$ is an $(n-k)$-bit vector produced by the PRNG and the right inverse of parity check matrix is computed through a public algorithm such as given in \cite{SobhiAfshar2009}.

Finally, the ciphertext is obtained as follows,
\begin{eqnarray}
c=(mG+e_P)P
\end{eqnarray}
\subsection{Decryption-Decoding}
We assume that the error vector $e$ is added to the ciphertext through a noisy channel between the transmitter and the receiver. Thus, we denote the received vector by
\begin{eqnarray}\label{r_vec}
r=c+e=(mG+e_p )P+e.
\end{eqnarray}

\begin{figure*}[!t]
	\centering
	\includegraphics[height=7cm]{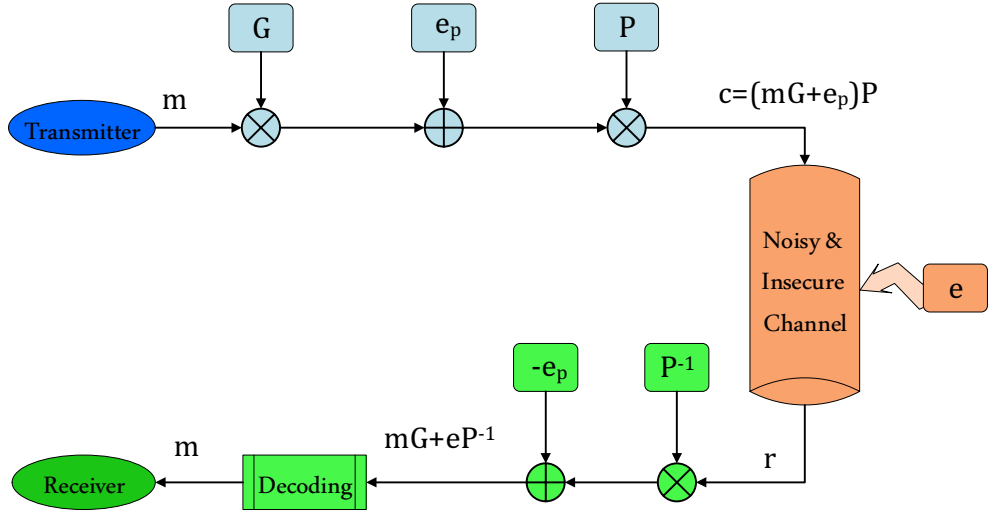}
	\caption{Block diagram of the proposed encryption-encoding / decryption-decoding scheme.}
	\label{fig_flowchart}
\end{figure*}

This algorithm works as follows:
\begin{enumerate}
	\item	Find the inverse permutation, $P^{-1}$.
	\item 	Multiply both sides of (\ref{r_vec}) by $P^{-1}$.
	\begin{eqnarray}\label{r_prime}
	r'=rP^{-1}=mG+e_p+eP^{-1}
	\end{eqnarray}
	\item	Subtract the perturbation vector $e_P$ from $r'$
	\begin{eqnarray}\label{c_prime}
	c'=r'-e_P=mG+eP^{-1}
	\end{eqnarray}
	\item	Decode $c'$ using a belief propagation algorithm to find $m$.
\end{enumerate}
Note that the $e'=eP^{-1}$ has the same Hamming weight as $e$.

Fig. \ref{fig_flowchart} shows block diagram of the joint encryption-encoding/decryption-decoding algorithms.
\subsection{The Code Parameters}\label{Code_param_subsec}
To deploy a fitting EG-QC-LDPC or PG-QC-LDPC code, length, rate, and density of the parity check matrix should be chosen properly. Our search results reflect the parameter values for different codes and we have summarized some suitable codes in Tables \ref{EG_codes} and \ref{PG_codes}. Although parameters of any particular future usage may need a value out of this range, we selected only those within the range of existing standards as samples.

\subsubsection{Code Rate}
The code rate of QC-LDPC codes with one block row is
\begin{eqnarray}
R=\frac{k}{n}=\frac{k_0p}{n_0p} = \frac{n_0-1}{n_0}.
\end{eqnarray}
In different communication standards, code rates vary from 1/5 in DVB-S2 \cite{etsi2015dvbs2} to 14/15 in IEEE 802.15.3c \cite{IEEE802.15.3c}. 
\subsubsection{Code Length}
The code lengths of EG-QC-LDPC and PG-QC-LDPC are as follows using the parameters of Tables \ref{table_EG} and \ref{table_PG}:
\begin{eqnarray}
n_{EG}=n_0 p_{EG}=n_0(q^m-1),
\end{eqnarray}
\begin{eqnarray}
n_{PG}=n_0 p_{PG}=n_0(\frac{q^{m+1}-1}{q-1}).
\end{eqnarray}
Similarly, code lengths in different standards bound our search for suitable parameters from 336 bits in ITU-T G9960 \cite{Itu-TG.9960std} to 64800 bits in DVB-S2 \cite{etsi2015dvbs2}.
\subsubsection{Parity Check Matrix Density}
A parity check matrix of density 0.01 or lower is categorized as a low density parity check matrix. LDPC codes of density about 0.001 have better error performance \cite{Baldi2013opt}. The density of the parity check matrices of EG-QC-LDPC and PG-QC-LDPC codes are given below, respectively,
\begin{eqnarray}
r_{EG}=\frac{\rho}{p}=\frac{q}{q^m-1}
\end{eqnarray}
\begin{eqnarray}
r_{PG}= \frac{\rho}{p}= \frac{q+1}{\frac{q^{m+1}-1}{q-1}}=\frac{q^2-1}{q^{m+1}-1}
\end{eqnarray}
where $\rho$ is the Hamming weight of the incident vector of each line in the geometry and $p$ is its length.

\section{Security and Performance}\label{S4Sec}
In order to evaluate a joint encryption-encoding scheme, also known as a secure channel coding scheme, we investigate the scheme from security and efficiency perspectives, namely, key size, error performance, and complexity of the scheme. Our goal in design of the FG-QC-LDPC joint encryption-encoding scheme is to decrease the key size as well as complexity of the scheme while improving error performance in comparison with the so far best previous schemes in the literature. In addition, keeping it secure against all known cryptanalytic attacks.
\subsection{Security}
Provable security for symmetric key cryptography is an open problem. There exists no natural hard problem that the security of the symmetric scheme can be reduced to. By the way, to examine symmetric schemes there is a method to reduce the security of that scheme to the problem of distinguishing between an oracle which encrypts a message with a random key and an oracle which outputs a random ciphertext \cite{Dent2006}. This reduction in oracle model for chosen-plaintext attack on symmetric key cryptosystems is given in \cite{bellare1997concrete}, which is applicable to analyze the security of modes of operations using a secure block cipher. Besides, Menezes in his talk argued against the role of provable security as a real ``proof". He claimed that provable security is a ``tool" and old-fashioned cryptanalysis is more reasonable in practical point of view \cite{Menezes2012prov}.

Based on the level of a priori knowledge, which is available to the cryptanalyst, there are different classes of cryptanalyses. We have examined our scheme against brute-force, ciphertext-only, message resend, and chosen-plaintext attacks. 

\subsubsection{Brute-Force Attack}
The secret key consists of the parity check matrix, H, the permutation matrix, P, and the seed of the PRNG, S. Each of these parameters must be chosen large enough in order to keep our scheme secure against brute-force attacks.

The number of parity check matrices of FG-QC-LDPC codes is
\begin{eqnarray}\label{N_FG}
N_{FG}=p^{n_0-1}\times \frac{\mathcal{N}_c!}{(\mathcal{N}_c-n_0)!},
\end{eqnarray}
where $\mathcal{N}_c$ is the number of cyclic classes on that geometry. Let $s$ denotes the desired security parameter of our scheme, then in the Eq.(\ref{N_FG}) we can simply assign the number of blocks, $n_0$, and the block size, $p$, in such a way that satisfy $N_{FG}>2^s$.
\begin{table*}[!t]
	\renewcommand{\arraystretch}{2.1}
	\caption{Parameters of EG-QC-LDPC Codes Designed for the Proposed Scheme}
	\label{EG_codes}
	\centering
	
	\begin{tabular}{|c|c|c|c|c|c|c|c|c|}
		\hline
		\bfseries \centering $n_0$ & \bfseries $q$ & \bfseries $m$& \bfseries $p$& \bfseries $\mathcal{N}_c$& \bfseries $n$& \bfseries $R$& \bfseries $r$& \bfseries $log_2(N_{EG})$\\[3pt]
		\hline\hline
		4 & 5 &6 &15624 &781 &62496	&0.75 &0.0003	&80.22 \\
		\hline
		5 & 3	&8	&6560	&1093	&32800	&0.80	&0.0005	&101.18 \\
		\hline
		6 & 2  &8	&255	&127	&1530	&0.8333	&0.0078	&81.73\\
		\hline 
		6 & 3 &6	&728	&121	&4368	&0.8333	&0.0041	&88.87 \\
		\hline 
		7 & 7  &4	&2400	&57	&16800	&0.8571	&0.0029	&107.65\\
		\hline 
		8 & 2  &9	&511	&255	&4088	&0.8750	&0.0039	&126.78 \\ 
		\hline
		9 & 2  &10	&1023	&511	&9207	&0.8889	&0.0020	&160.86\\
		\hline
		15 & 2  &10	&1023	&511	&15345	&0.9333	&0.0020	&274.64\\
		\hline
	\end{tabular}
\end{table*}
\begin{table*}[!t]
	\renewcommand{\arraystretch}{2.1}
	\caption{Parameters of PG-QC-LDPC Codes Designed for the Proposed Scheme}
	\label{PG_codes}
	\centering
	
	\begin{tabular}{|c|c|c|c|c|c|c|c|c|}
		\hline
		\bfseries \centering $n_0$ & \bfseries $q$ & \bfseries $m$& \bfseries $p$& \bfseries $\mathcal{N}_c$& \bfseries $n$& \bfseries $R$& \bfseries $r$& \bfseries $log_2(N_{PG})$\\[3pt]
		\hline\hline
		5	&4	&6	&5461	&273	&27305	&0.8000	&0.0009	&90.07 \\
		\hline
		6	&2	&8	&511	&85	&3066	&0.8333	&0.0059	&83.18  \\
		\hline
		6	&2	&9	&1023	&170	&6138	&0.8333	&0.0029	&94.32 \\
		\hline 
		8	&3	&6	&1093	&91	&8744	&0.8750	&0.0037	&122.26  \\
		\hline 
		11	&2	&8	&511	&85	&5621	&0.9091	&0.0059	&159.50  \\
		\hline 
		13	&5	&5	&3906	&130	&50778	&0.9231	&0.0015	&233.57  \\ 
		\hline
		15	&3	&7	&3280	&273	&49200	&0.9333	&0.0012	&284.34  \\
		\hline
	\end{tabular}
\end{table*}
Considering this constraint as well as those described in Section \ref{Code_param_subsec}, our search for parameters of the suitable codes resulted various examples summarized in Tables \ref{EG_codes} and \ref{PG_codes}.

For the permutation matrix, its block size, $l$, must be chosen as great as $l!>2^s$ condition holds. With this in mind, the block length can be kept considerably short. Giving an example, if $l\geq33$ then $l!>2^{120}$.

For the PRNG, the length of the seed must simply be chosen larger than the security parameter, $s$.

\subsubsection{Ciphertext-Only Attack}
The goal of this attack is to recover the plaintext from its ciphertext without any knowledge of the key. In code-based cryptosystems, this is interpreted as decoding a coded message without access to its parity check or generator matrices. To achieve this goal, the adversary needs to solve the \textit{general decoding problem}, which is known to be NP-hard problem \cite{Berlekamp1978}. This cryptanalysis was applied on the McEliece-like public-key code-based cryptosystems \cite{LeeBrickell1988,Becker2012,May1817}, whose public generator matrix is algebraically equivalent to their secret key generator matrix. Since in our symmetric key code-based schemes the parity-check matrix is kept secret, this attack is not feasible on our scheme in polynomial time.

\subsubsection{Message Resend Attack}
The aim of the message resend attack is to find the perturbation vector, $e_P$, used by the transmitter and then recover the message in the following manner. Suppose that the transmitter sends $c_1=(mG+e_{P1} )P$ to the receiver. The attacker, as the man in the middle, alter some bits of $c_1$ such that the receiver receives a false or undecodable vector. Therefore, the receiver has to make a request to the transmitter for resending the message, $m$. This time, the transmitter encrypts the same message using a different perturbation vector $e_{P2}$, as $c_2=(mG+e_{P2})P$. This scenario is called message resend \cite{berson1997}. In this situation, the attacker has access to two different ciphertexts $c_1$ and $c_2$ of the same message $m$. So the attacker can obtain the following equation and thereby guessing the positions of non-zero entries of $e_{P1}$ and $e_{P2}$.
\begin{eqnarray}
c_1+c_2=(e_{P1}+e_{P2})P
\end{eqnarray}

This attack is only feasible when the perturbation vectors have low Hamming weight. Since the used perturbation vectors in the proposed scheme are generated uniformly at random, it is not feasible to find each of $e_{P1}$  and $e_{P2}$ from $c_1+c_2$. Moreover, the secret permutation matrix, $P$, changes the location of ‘1’s and ‘0’s.

Apart from these issues while applying this attack, error correction capability of capacity approaching FG-QC-LDPC codes could obviate the need for resending the message. Because the alterations made by the attacker can be recovered by the error correction code. Thus, the message resend scenario does not occur.

\subsubsection{Chosen-Plaintext Attack}

There are two major chosen-plaintext attack scenarios against secret key code-based cryptosystems, namely Struik-Tilburg \cite{StruikTilburg1987} and Rao-Nam \cite{RaoNam1989} attacks.

\paragraph{Struick-Tilburg Attack}
Struik and Tilburg \cite{StruikTilburg1987} proposed a chosen-plaintext attack against secret-key code-based cryptosystems. In this attack two plaintexts $m_1$ and $m_2$ are chosen in a way that they are only different on their $i^{th}$ position, i.e. $m_1-m_2=u_i$. As a result, the corresponding ciphertext difference is
\begin{eqnarray}
c_1-c_2=u_i GP+(e_{P1}-e_{P2})P=g'_i+(e_{P1}-e_{P2} )P,
\end{eqnarray}
where $g'_i$ is the $i^{th}$ row of the generator matrix $G'=GP$. The attacker will repeat the procedure for the same $u_i$ and different perturbation vectors $e_p$ until a set of all possible ciphertexts of $u_i$ are collected. The cardinality of this set is equal to the number of total perturbation vectors, $N_e$. Doing a brute-force over all perturbation vectors, $g'_i$ is obtained. Repeating the above scenario for all $i$ reveals the whole matrix $G'$.

The work factor of this attack is of $\Omega(knN^2_e log_2⁡(N_e))$ \cite{RaoNam1989}. Therefore, this attack will be successful only if the set of all perturbation vectors, $N_e$, has small cardinality. In FG-QC-LDPC joint encryption-encoding scheme $N_e=2^{(n-k)}=2^{p}$ and according to Tables \ref{EG_codes} and \ref{PG_codes}, $min(p)=255$. So the work factor of this attack is of $\Omega(1275\times 1530\times 2^{510} \times 255)$, which is dramatically large and therefore the Struik-Tilburg attack is not applicable to the FG-QC-LDPC joint encryption-encoding scheme in polynomial time.

\paragraph{Rao-Nam Attack}
Rao and Nam \cite{RaoNam1989} proposed their attack based on the previously mentioned Struik-Tilburg \cite{StruikTilburg1987} attack. They similarly used chosen-plaintexts $m_1$ and $m_2$ differing only in one position. They noticed that when the perturbation vectors has low Hamming distance the the attacker can use majority voting to estimate $g'_i$ and thereby revealing the whole matrix $G'$. The work factor of this attack, obtained by Rao and Nam \cite{RaoNam1989}, is $\Omega(N_e^k)$. Based on Table \ref{EG_codes} and \ref{PG_codes}, The minimum work factor of this attack on our proposed scheme is $\Omega(255^{1275})$. Therefore, the FG-QC-LDPC joint scheme is far more secure to be threatened by this attack.
\subsection{Key Size}
The secret key of the FG-QC-LDPC joint encryption-encoding scheme as mentioned in Section \ref{Key_gen} consists of the seed vector for a PRNG ($S$), the parity check matrix ($H$), and the permutation matrix ($P$).
\begin{eqnarray}
|K_s |=|S|+|K_P |+|K_H|
\end{eqnarray}

First, choosing a suitable PRNG for each application, keeps the size of the seed at a desirable extent. Comparing PRNGs is not in the scope of this paper. However, as pointed in Section \ref{S1Intro}, it is not recommended to use simple LFSRs based on the reasons mentioned in \cite{Lee2017}. In our example, we use Sosemanuk-128 stream cipher as an example of PRNG \cite{Berbain2008}. The size of the seed vector of this PRNG is only 128 bits.

Owing to quasi-cyclic structure of the parity check matrix, storing only the first row of each circulant block of this matrix suffices to create the parity check matrix. Furthermore, thanks to the finite geometry construction of these blocks, the whole first row of each block can be produced by only knowing the location of two ``1"s on each. Since each row is an incident vector of a line on finite geometry, the two ``1"s indicate the two points where a line go through them. Thus, these two location numbers can regenerate the line and its incident vector.

We introduce a practical method to achieve the information theoretic lower bound for storing the first row of each circulant block. In this regard, we need to identify two things, the representative of cyclic class and the number of cyclic right shift to obtain the first row. The following constraints must be considered to assign a unique line as a representative for each cyclic class.

\begin{enumerate}[i)]
	\item \label{i}The first element of its incident vector must be ``1".
	\item \label{ii}The next ``1" in the incident vector must be located at the nearest possible locations among all lines of the class.
\end{enumerate}
If the non-zero elements of the incident vector of the representative are $\alpha^{j_1},\alpha^{j_2},\dots ,\alpha^{j_\rho}$  , (i) forces that $j_1=0$  and (ii) forces $j_2-j_1<\min{(j_{i+1}-j_i) ,(j_1-j_\rho)(mod\: p)}$. By using this method we only store $j_2$ to indicate the cyclic class. This needs only $\lceil \log_2({\frac{p}{\rho}})  \rceil $ bits. Another $\lceil \log_2({p})  \rceil $ bits is needed to indicate the amount of cyclic shift for the first row of each block. As a result, the amount of memory to store each circulant block of FG-QC-LDPC parity check matrix is
\begin{eqnarray}
\lceil \log_2({\frac{p}{\rho}})  \rceil + \lceil \log_2({p})  \rceil & bits.
\end{eqnarray}
While a permutation in the rows of the parity check matrix makes no difference in the code, we can suppose that the first circulant block (or one of the others) made by the representative without being cyclically shifted. As a result, the whole parity check matrix with one block row and $n_0$ blocks needs the following amount of memory to be stored. 
\begin{eqnarray}
|K_H|= n_0 \times \lceil \log_2({\frac{p}{\rho}})  \rceil + (n_0-1) \times \lceil \log_2({p})  \rceil & bits
\end{eqnarray}

The block diagonal permutation matrix, $P$, in this scheme will be stored in similar way as in \cite{Barbero2000}. The size of the permutation matrix of the key is as follows. Where $l$ is the length of each block and $l'=2^{\lfloor\log_2(l)\rfloor}$.
\begin{eqnarray}
|K_P|= l(\log_2 l' +1) -2l'+1
\end{eqnarray}

Tables \ref{Key80} and \ref{Key120} show examples of codes and their key sizes for 80 bits and 120 bits security parameters, respectively.
\begin{table}[!t]
	\renewcommand{\arraystretch}{1.8}
	\caption{The key size of the proposed system with 80 bits security parameter}
	\label{Key80}
	\centering
	
	\begin{tabular}{|C|D|H|W|}
		\hline
		\bfseries \centering Secret key parts & \bfseries Proposed parameters & \bfseries Computational complexity to find key & \bfseries Key size\\[3pt]
		\hline\hline
		H	& $q=3, m=6, n_0=6$	& $2^{88.87}$	&92 bits \\
		\hline
		P	& $l=26$	& $2^{88.38}$	&99	bits  \\
		\hline
		S	&Sosemanuk-128	& $>2^{80}$	&128 bits\\
		\hline 
		\multicolumn{3}{|c|}{Total Key size}	&\bfseries 319 bits	 \\
		\hline
	\end{tabular}
\end{table}

\begin{table}[!t]
	\renewcommand{\arraystretch}{1.8}
	\caption{The key size of the proposed system with 120 bits security parameter}
	\label{Key120}
	\centering
	
	\begin{tabular}{|C|D|H|W|}
		\hline
		\bfseries \centering Secret key parts & \bfseries Proposed parameters & \bfseries Computational complexity to find key & \bfseries Key size\\[3pt]
		\hline\hline
		H	& $q=3, m=6, n_0=8$	& $2^{121.56}$	&126 bits \\
		\hline
		P	& $l=52$	& $2^{225.58}$	&249	bits  \\
		\hline
		S	&Sosemanuk-128	& $>2^{120}$	&128 bits\\
		\hline 
		\multicolumn{3}{|c|}{Total Key size}	&\bfseries 503 bits	 \\
		\hline
	\end{tabular}
\end{table}

The key size of the proposed scheme, taking the advantages of FG-QC-LDPC codes, decreases to 319 bits where the key size of the last known similar system was 2272 bits \cite{Esmaeili2015}. Table \ref{Key_size} compares the key size of the proposed system with those known similar systems.

\begin{table}[!t]
	\renewcommand{\arraystretch}{1.8}
	\caption{Comparison of the key size of the proposed scheme with the preceding schemes}
	\label{Key_size}
	\centering
	
	\begin{tabular}{|D|M{3.5cm}|C|}
		\hline
		\bfseries \centering Cryptosystem  & \bfseries Code & \bfseries Key size \\[3pt]
		\hline\hline
		Rao\cite{Rao1984}	& C(1024,524)&	2 Mbits \\
		\hline
		Rao-Nam\cite{RaoNam1989}	&C(72,64)	&18 Kbits \\
		\hline
		Struik-Tilburg\cite{StruikTilburg1987}	&C(72,64)	&18 Kbits\\
		\hline 
		Barbero-Ytrehus\cite{Barbero2000}	&C(30,20) over $GF(2^8)$	&4.9 Kbits	 \\
		\hline
		Sobhi Afshar, et al.\cite{SobhiAfshar2009}	&C(2044,1024)	&2.5 Kbits\\
		\hline
		Hooshmand, et al.\cite{hooshmand2015improving}	&C(2470,2223)	&3.55 Kbits\\
		\hline
		Esmaeili, et al.\cite{Esmaeili2014}	&C(2048,1536)	&2191 bits\\
		\hline
		Esmaeili, Gulliver\cite{Esmaeili2015}&	C(2048,1536)&	2272 bits\\
		\hline
		Mafakheri, et al.\cite{mafakheri2017}&	C(2048,1781)&	1611 bits\\
		\hline
		The proposed scheme&	C(4368,3640)	&319 bits\\
		\hline
		
	\end{tabular}
\end{table}
\subsection{Error Performance}
\begin{figure*}[!t]
	\centering
	\includegraphics[height=7cm]{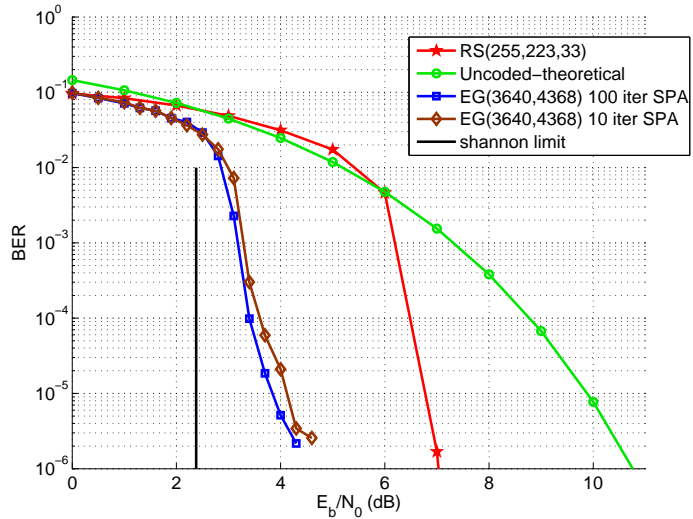}
	\caption{Performance comparison of log-SPA decoder of 10 and 100 iterations with Reed-Solomon code.}
	\label{ber}
\end{figure*}

At the receiver the FG-QC-LDPC code used in our system is decoded by a logarithmic Sum-Product decoder. We took the following considerations to simulate encoding, channel, and decoding processes. In our simulation codewords transmitted via a Quadratic Phase Shift Keying (QPSK) channel with additive white Gaussian noise. The receiver has access to soft information of channel. We compared decoders of 10 and 100 iterations with a Reed-Solomon code in Fig. \ref{ber}. This figure shows that there is no remarkable improvement in 100 iterations decoding in compare to 10 iterations.

\subsection{Complexity}
There are two separate process which their computational complexity needs to be assessed, encryption-encoding and decryption-decoding processes.
\subsubsection{Encryption-Encoding}
The complexity of this process can be calculated as follows.
\begin{eqnarray}\label{C_enc1}
\mathcal{C}_{Enc}=\mathcal{C}_{mul}(mG)+\mathcal{C}_{add}(mG+e_P)+\\
\mathcal{C}_{mul}(H^{-1}.s)+\mathcal{C}_{mul}(P)\nonumber
\end{eqnarray}

In this equation $\mathcal{C}_{add} (mG+e_P)$ stands for adding two $n$-bit vectors which consume $n$ binary operations. Multiplying a vector by a sparse matrix $a_{1×n}×B_{n×n}$, needs $nw$ binary operations \cite{Baldi2008}, where $w$ is the Hamming weight of rows of the sparse matrix. Here permutation matrix, $P$, has $w=1$ so $\mathcal{C}_{mul}(P)=n$.

The generator matrix, $G$, and the inverse of parity check matrix, $H^{-1}$, are dense matrices and need $kn$ and $(n-k)n$ binary operations respectively. By the way, their quasi cyclic property leads to a 92\% lower computational complexity in multiplying operations \cite{Baldi2008}. Therefor, we can conclude that:
\begin{eqnarray}\label{C_enc2}
\mathcal{C}_{Enc}=0.08\times k.n+n+0.08\times (n-k).n+n\\
=\frac{0.08n+2}{R}  \nonumber,
\end{eqnarray}
where $R=\frac{k}{n}$.

\subsubsection{Decryption-Decoding}
The complexity of this process is obtained as follow.
\begin{eqnarray}\label{C_dec1}
\mathcal{C}_{Dec}=\mathcal{C}_{mul}(r\times P^{-1})+\mathcal{C}_{add}(r'+e_P)\\
\mathcal{C}_{mul}(H^{-1}s)+ \mathcal{C}_{mul}(c'\times H^T)+ \mathcal{C}_{SPA}\nonumber
\end{eqnarray}

The complexity of the Sum-Product Algorithm, as mentioned in \cite{Baldi2008}, is
\begin{eqnarray}\label{C_SPA}
\mathcal{C}_{SPA}=I_{avg}.n[d(8\rho+12R-11)+\rho].
\end{eqnarray}
In this equation $I_{avg}$ is the average number of decoding iterations and $d$ is the number of quantization bits in analog-to-digital converter. Finally letting $I_{avg}=10$ and $d=6$, the number of binary operations for each information bit to be decrypted-decoded is
\begin{eqnarray}\label{C_deck}
\mathcal{C}_{Dec/k}=\frac{1}{R}(2+n-k+n_0\rho+490\rho+720R-110).
\end{eqnarray}
In this equation it is obvious that the complexity of decryption-decoding algorithm is linearly proportional to the redundancy $(n-k)$. 
\section{Conclusion}\label{S5con}
This paper introduces a joint encryption-encoding scheme, also known as secure channel coding, using QC-LDPC codes based on finite geometry. We have taken advantage of FG-QC-LDPC codes to shorten the secret key to 20\% of that of the previously best known similar systems.

Thanks to the LDPC codes and their fast iterative decoding, the error performance of the proposed scheme is among the best of the literature. The FG-QC-LDPC joint encryption-encoding scheme is secure against cryptanalyses of such cryptosystems.

The joint algorithm leads to lower complexity than conventional encryption-then-encoding methods. We have shown that our system can provide reliability and security simultaneously with the lower cost of one joint system rather than two disjoint systems.
\section*{Acknowledgment}

The authors would like to thank anonymous reviewers
for their precious and productive comments.

\bibliographystyle{splncs04e}

\bibliography{MyCollectionfe}

\end{document}